\documentclass[%
 reprint,
 amsmath,amssymb,
 aps,
]{revtex4-2}

\usepackage{graphicx}
\usepackage{dcolumn}
\usepackage{bm}

\begin{document}

\preprint{APS/123-QED}

\title{Pulsating dissipative solitons in a Mamyshev oscillator}
\author{Bo Cao}

\author{Kangjun Zhao}%
\author{Chenxin Gao}%

\author{Chengying Bao}%
 \altaffiliation{cbao@tsinghua.edu.cn}
\author{Changxi Yang}%
 \email{cxyang@tsinghua.edu.cn}
\affiliation{State Key Laboratory of Precision Measurement Technology and Instruments, Department of Precision Instruments, Tsinghua University, Beijing 100084, China}

\author{Xiaosheng Xiao}%
\affiliation{State Key Laboratory of Information Photonics and Optical Communications, School of Electronic Engineering, Beijing University of Posts and Telecommunications, Beijing 100876, China}

\date{\today}

\begin{abstract}
Mode-locked fiber lasers provide a versatile playground to study dissipative soliton (DS) dynamics. The corresponding studies not only give insights into soliton dynamics in dissipative systems, but also contribute to femtosecond fiber laser design. Recently, Mamyshev oscillators (MOs), which rely upon a pair of narrow filters with offset passing frequencies, have emerged as a promising candidate for high power, ultrabroad bandwidth pulse generation. 
To date, only stable mode-locking states in MOs have been reported. 
Here, we present a comprehensive experimental and numerical investigation of pulsating DSs in an ytterbium MO. By reducing the filter separation down to 4 nm, we observe pulsation in a single pulse state as well as a soliton molecule state. In the single pulse state, the output pulse energy can vary as large as 40 times in our MO. Single-shot spectra measured by the dispersive Fourier transform (DFT) method reveal the spectral bandwidth breathing during pulsation and enables the observation of soliton explosion in a pulsation state. In addition, pulsation with a period lasting 9 round trips and even a chaotic pulsation state are also observed. Numerical simulations based on a lumped model qualitatively agree with our observation. Our results enrich DS dynamics in MOs and show the impact of filter separation on the stability of MOs.

\end{abstract}

\maketitle


\section{Introduction}

Passively mode-locked fiber lasers can support dissipative solitons (DSs) via the delicate balance between nonlinearity, dispersion, gain and loss \cite{grelu2012dissipative}. The stability endowed by DS formation makes mode-locked fiber lasers invaluable tools for applications spanning from nonlinear microscopy \cite{Wise_NP2013}, precise timing dissemination \cite{Kartner_LPR2008} to optical frequency combs \cite{Diddams_Science2020optical}. DS dynamics in mode-locked lasers can be governed by the cubic-quintic Ginzburg-Landau equation (CGLE) \cite{grelu2012dissipative}. In addition to stable DSs, pulsating DSs, where DSs vary periodically in propagation, is also an eigenstate \cite{deissler1994periodic,soto2000pulsating,akhmediev2001pulsating,spaulding2002nonlinear}. Extreme pulse energy variation exceeding two orders of magnitude has been theoretically predicted for pulsating DSs \cite{chang2015extreme}. The CGLE averages the DS propagation within a single round trip. When considering the round trip nature of lasers, pulsation of DSs manifests as period 
multiplication (period-$N$ pulsation means DSs only repeat after $N$ round trips) in mode-locked fiber lasers \cite{soto2004bifurcations,zhao2004observation,zhao2004pulse,zhao2007period,wang2017subsideband}. 
Due to the bandwidth limitations of electronic digitizers, typically only pulse energy variation was measured in early reports onDS  pulsation and it was quite challenging to measure fast pulsation dynamics regarding other parameters in real-time. With the emergence of the dispersive Fourier transform (DFT) method \cite{goda2013dispersive}, single-shot spectra during pulsation have been measured recently \cite{liu2020visualizing,chen2019dynamical,peng2019breathing,peng2019experimental}. These measurements reveal that pulsating DSs can be associated with strong spectral bandwidth breathing \cite{peng2019breathing} or even invisible if only time domain pulse train is measured \cite{liu2020visualizing}.

To date, pulsating DSs have been observed in fiber lasers  mode-locked by material-based saturable absorbers \cite{wang2018self,chen2019dynamical,liu2020visualizing} and by nonlinear polarization rotation (NPR) \cite{zhao2004observation,zhao2004pulse,zhao2007period,wang2017subsideband,peng2019breathing}. Recently, a new saturable absorption mechanism based on the Mamyshev regenerators \cite{mamyshev1998all} have been leveraged for DS mode-locking in fiber lasers \cite{regelskis2015ytterbium,liu2017megawatt,sidorenko2018self,liu2019femtosecond,repgen2020mode,boulanger2020all}. This effective saturable absorber relies upon spectral broadening induced by self-phase modulation (SPM) and a pair of filters with offset transmission frequencies  \cite{mamyshev1998all}. Thus, high power optical field will pass the filter pair with higher transmission than low power field, resulting in an effective saturable absorber \cite{rochette2008multiwavelength,pitois2008generation,north2014regenerative}. Mode-locked fiber lasers based on this mechanism, referred to as Mamyshev oscillators (MOs), show great prospects in reaching high output pulse energy \cite{regelskis2015ytterbium,liu2017megawatt,liu2019femtosecond}. For example, a MO delivering 50 nJ pulses with a peak power exceeding 10 MW has been reported \cite{liu2019femtosecond}. Moreover, MOs can enable the generation of ultrabroad spectra and few-cycle pulses. Ma et al. have demonstrated a MO with an output spectrum spanning $\sim$400 nm that can be externally dechirped to $\sim$20 fs \cite{ma2020ultrabroadband}.

Besides high energy ultrashort pulse generation, the unique architecture of MOs makes them a new playground to study DS dynamics. Indeed, multi-soliton states including harmonic mode-locking \cite{wang2019pattern,poeydebat2020all,yan2021pulse} 
and soliton molecule \cite{xu2020multipulse} have been observed in MOs in addition to stable single-pulse states. However, these DSs remain stable and pulsating DSs in MOs have not been reported to our knowledge.

Here, we report experimental and numerical observation of pulsating DSs in a MO operating around 1030 nm. Pulsating DSs are observed by controlling the pump power as well as the polarization state under a relatively small filter separation (4 nm). Single-shot spectra of these pulsating DSs are also measured by the DFT method, which indicate that our MO admits various pulsation dynamics. Spectral breathing associated with pulse energy pulsation occurs for both a single pulse state or a soliton molecule state in period-2 pulsation 
states. In the single pulse state, pulsating DSs whose output energy varies as large as 40 times during pulsation are observed. Period-3 pulsation 
is observed for two wide spaced DSs pulsating in different ways. Soliton explosion \cite{cundiff2002experimental,runge2015observation,peng2019experimental,chen2019dynamical,zhou2020breathing} is also observed in this state, which constitutes the first observation of soliton explosion in MOs, to our knowledge. 
In addition to these short period pulsation, pulsation lasting a relatively long period (e.g., 9 round trips) and even chaotic pulsation (no regular period) also exist in the MO.
Numerical simulations based on a lumped model including the polarization dynamics are in reasonably qualitative agreement with the experimental observation and verify the existence of various pulsation dynamics in MOs. Our observation not only adds to the possible DS states in MOs but also aids the design of MOs. 

\section{Laser design and stable operation}

\begin{figure}[t]
\centering\includegraphics[width=\linewidth]{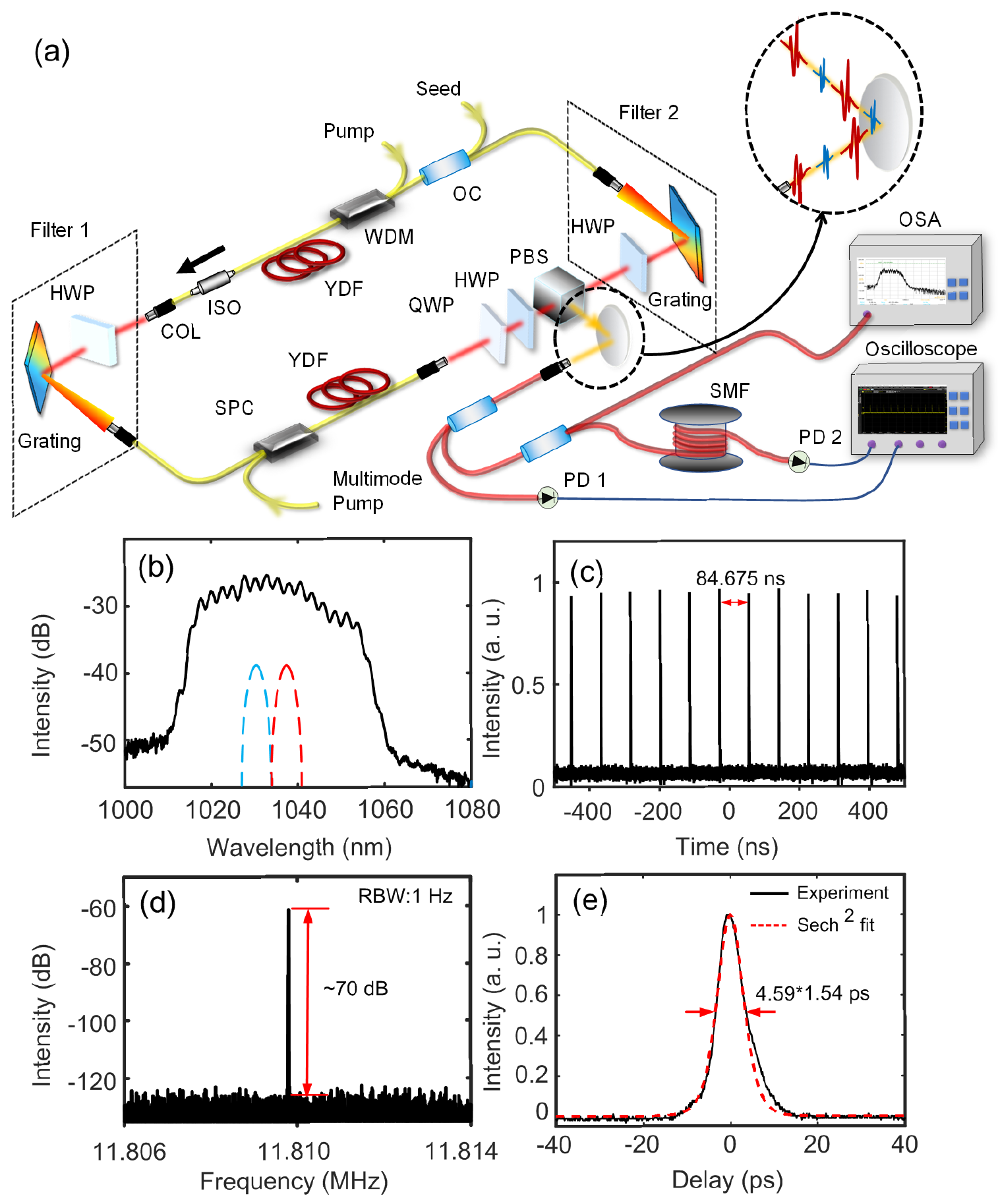}
\caption{ \textbf{Experimental setup and laser output in a stable state.} (a) Schematic of the Mamyshev oscillator (MO) and the measurement system. The top inset illustrates a train of dissipative solitons in pulsation. OC: optical coupler, WDM: wavelength division multiplexer, COL: collimator, ISO: isolator HWP: half waveplate, QWP: quater waveplate, PBS: polatization beam splitter, SMF: single mode fiber, SPC: signal-pump combiner, PD: photodetector. (b) Optical spectrum of the MO outoput with a filter separation of 7 nm. Dashed blue and red curves correspond to the transmission of the filters. (c) Pulse train of the MO output; (d) RF spectrum; (e) autocorrelation trace, showing a pulse width of 4.6 ps.}
\label{fig1}
\end{figure}

\begin{figure*}[t]
\centering{\includegraphics[width=\linewidth]{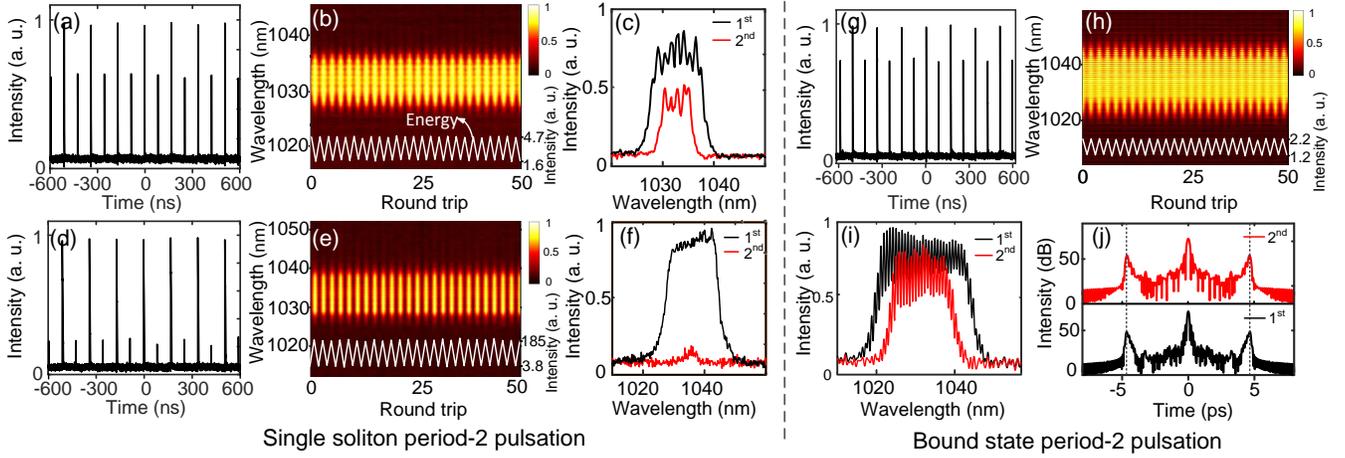}}
\caption{\textbf{Single soliton and soliton molecule pulsation.} (a) and (d) Pulsation pulse trains of the two typical operations. (b) and (e) DFT measured single-shot spectra corresponding to (a) and (d), respectively. The white lines are the integrated pulse energy for the pulsating DS. (c) and (f) Single-shot spectra in two successive round trips. (g) Pulsating pulse train of a soliton molecule. (h) Single-shot spectra of the pulsation state. (i) Spectra of two successive round trips. The spectra show modulation due to the interference between two closely spaced pulses. (j) Fourier transform of panel (i). Peaks at 4.65 ps illustrates the pulse separation of the soliton molecule.}
\label{fig2}
\end{figure*}

\begin{figure*}[ t]
\centering\includegraphics[width=\linewidth]{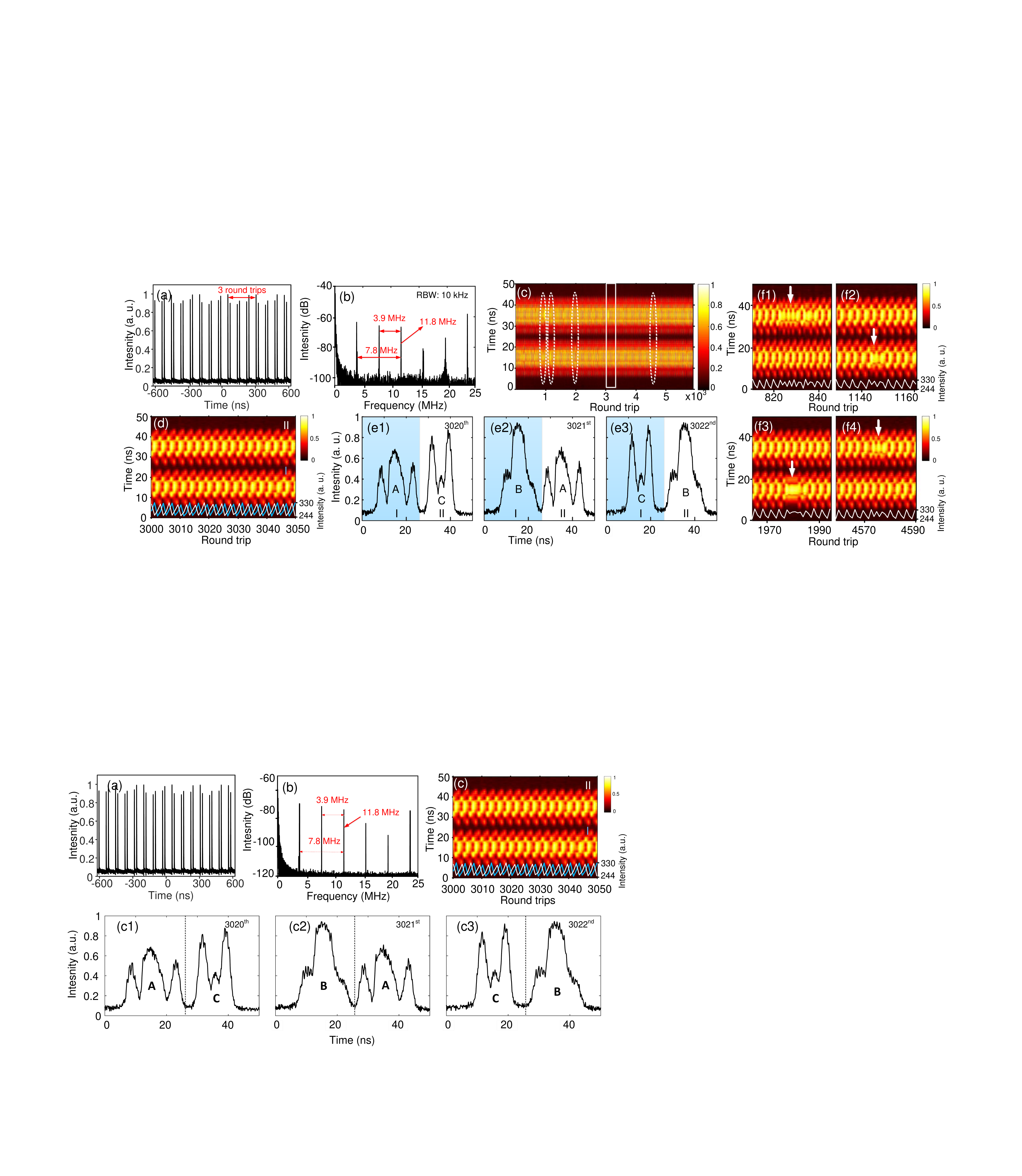}
\caption{\textbf{Period-3 pulsation and explosion for two wide spaced solitons.} (a) The soliton trains showing pulse energy during pulsation. (b) RF spectrum of the pulse train exhibiting additional RF tones besides the fundamental repetition rate 11.8 MHz. (c) Single-shot spectra in 6000 round trips. Note that the y-axis of this panel is labeled as time rather as wavelength derived from DFT (i.e., time to wavelength mapping). This is because the two pulses are plotted with increasing time but they operate around the same wavelength. (d) Zoom in of the data marked by the rectangular shape in panel (c) showing a regular period-3 pulsation. The light blue and white curves are the pulse energy change for dissipative soliton \textrm{I} and \textrm{II}, respectively. (e1)-(e3) Spectra of the pulsating dissipative solitons within a single pulsation period. Dissipative soliton \textrm{I} and \textrm{II} both undergo spectra A, B, C in pulsation, but experience them in different orders. (f1)-(f4) Zoom in of the data marked by the ellipse shape in panel (c). The single-shot spectra show abrupt spectral collapse in soliton explosion, as indicated by the arrows. The while line shows the integrated pulse energy for soliton in explosion.}
\label{fig3}
\end{figure*}

\begin{figure}[t]
\centering\includegraphics[width=\linewidth]{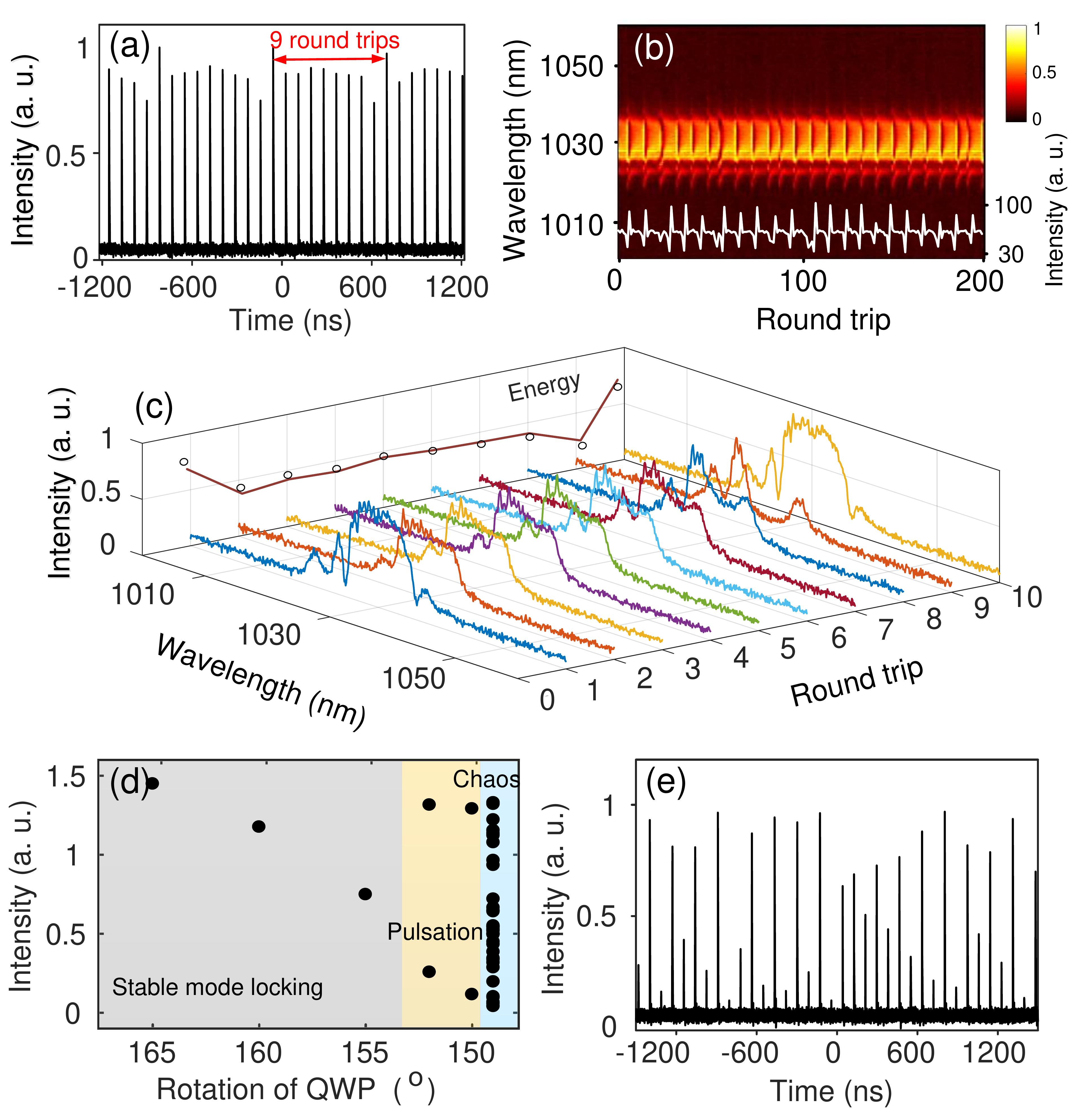}
\caption{\textbf{Long-period pulsation and bifurcation to a chaotic state.} (a) A soliton train showing period-9 pulsation. (b) Single-shot spectra of the period-9 pulsation state. The white line shows the integrated  during pulsation. (c) Details of the spectral evolution within one pulsation period. (d) Pulse energy bifurcation diagram when rotating the quarter waveplate. (e) Pulse train in a chaotic state under an waveplate orientation of 148$^\circ$.}
\label{fig4}
\end{figure}

\begin{figure*}[t]
\centering\includegraphics[width=\linewidth]{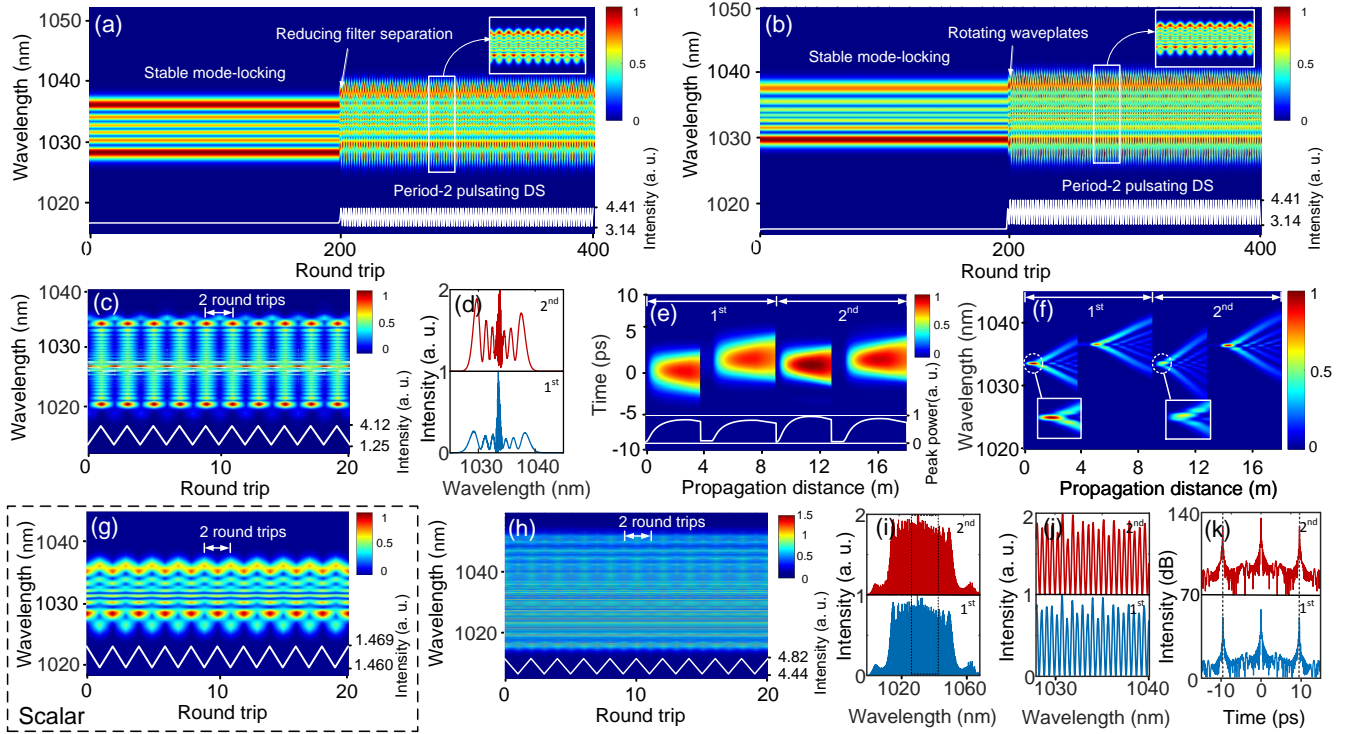}
\caption{\textbf{Simulation of the pulsating DSs in a vector model.} (a) Transition from stable mode-locking to pulsating DS by  reducing the filter separation. (b) Transition into pulsation by rotating the waveplates. The white curves in panels (a, b) show the integrated  change. (c) Spectral dynamics of a period-2 pulsation state with a relatively large pulse energy change. (d) Single-shot spectra of the period-2 pulsating DS in panel (c). (e) Temporal and (f) spectral dynamics within two successive cavity round trips for the pulsation state in panel (c). The white solid line in panel (e) is the peak power of the pulse in the cavity, while the insets of panel (f) depict the zoom in spectra right after the filter 1 of two successive round trips. (g) Period-2 pulsating DS simulated by a scalar model (the other panels are all based on the vector model). The white curve is the integrated  showing a very small change in pulsation. (h) Evolution of spectrum of a soliton molecule in period-2 pulsation. (i) Single-shot spectra of the period-2 pulsating bound soliton state in panel (g). (j) Zoom in of the data in panel (i). (k) Fourier transform of panel (i) shows the pulse separation remains unchanged in pulsation.}
\label{fig5}
\end{figure*}

\begin{figure*}[t]
\centering\includegraphics[width=\linewidth]{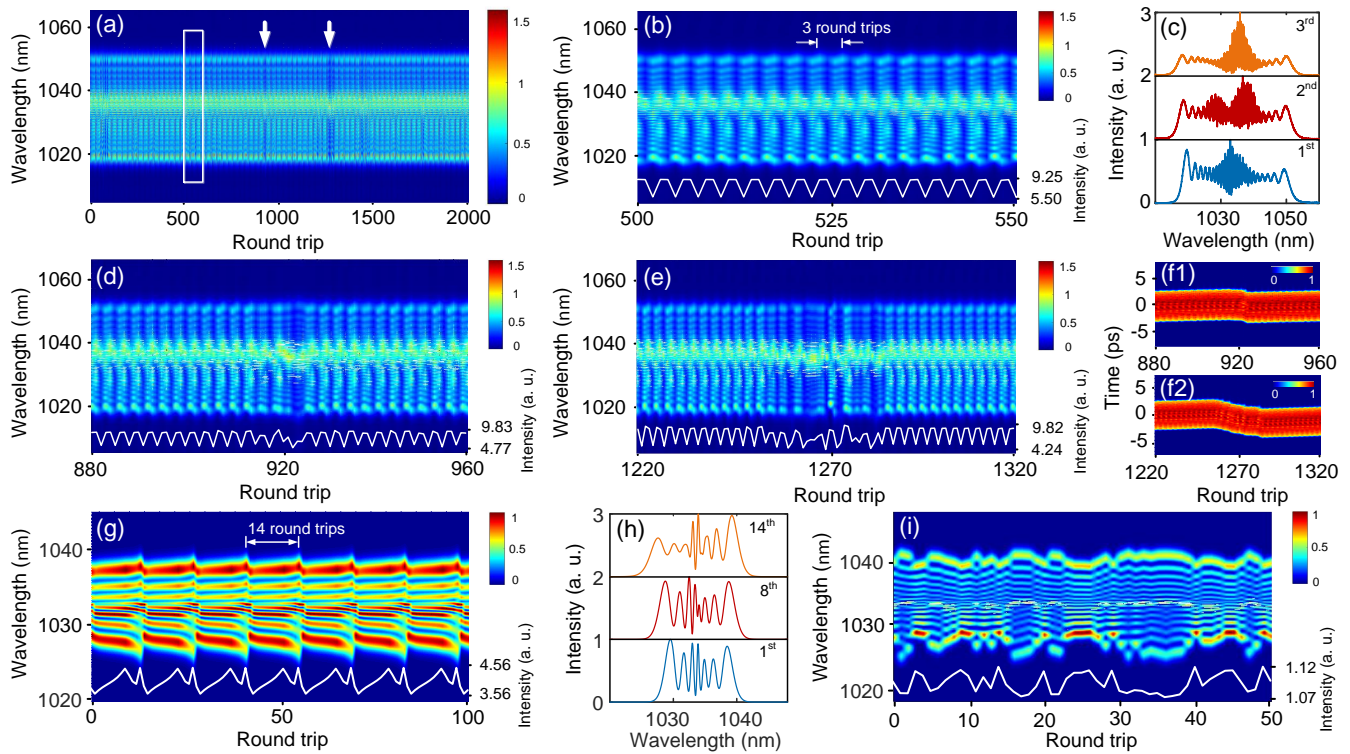}
\caption{\textbf{Simulations of the pulsating soliton explosion and the long-period pulsation.} (a) Evolution of the spectra over 2000 round trips in a period-3 pulsation with soliton explosion, marked by the arrows. (b) Zoom in of the data marked by the rectangular shape in panel (a), showing regular period-3 pulsation. The white curve shows the integrated . (c) Single-shot spectra of the period-3 pulsating DS. (d) and (e) Zoom in of the data marked by the arrows in panel (a). The white curves show the pulse energy change during explosion. (f1) and (f2) Temporal shift of the pulsating soliton explosion depicted in panels (d) and (e), respectively. (g) Evolution of the spectra of a long-period (period-14) pulsation state. The white curve stands for the pulse energy change. (h) Three representative single-shot spectra within a pulsation period shown in panel (g). (i) Spectral evolution of a chaotic pulsation state; the corresponding pulse energy (white curve) evolution loses periodicity. }
\label{fig6}
\end{figure*}

Scheme of the experimental setup is shown in Fig. \ref{fig1}(a). The laser cavity consists of two Mamyshev regenerators. The upper arm comprises a 2 m core pumped Yb-fiber with a small core diameter of 6 $\mu$m so as to enhance the nonlinearity. Thus, a 976 nm single mode pump is sufficient to amplify the seeding pulses to pass filter 2 via SPM induced spectral broadening (see Fig. \ref{fig1}(a)). For the second arm, a 1.5 m double-cladding Yb-fiber (LMA-10/125-YDF, Nufern) is pumped by a 976 nm multimode pump so that a higher output power is possible. All the remaining passive fibers in the cavity are HI1060 fibers. An isolator is used to ensure unidirectional propagation. To realize offset filtering, a pair of 600 lines/mm reflective diffraction gratings (GR13-0610, Thorlabs) are used to form two filters whose passing band can be adjusted by the grating orientation and the collimator position (dotted squares in Fig. \ref{fig1}(a)) \cite{liu2017megawatt}. Two half waveplates are placed before the gratings to maximize the diffraction efficiency. The filter bandwidth is measured to be $\sim$2 nm. The other waveplates and a polarizing beam splitter (PBS) are used to create a NPR-like transmission in the cavity that can assist initialization of the mode-locking. Note that the waveplates can be adjusted over a wide range (in some cases 360$^\circ$ for the QWP) without losing mode-locking after pulses generated, which elucidates the effect of Mamyshev regenerators on stabilizing the mode-locked pulses.

The pulses are sampled by the PBS as output. The pulse train is detected by a 5 GHz photodetector (PD 1) and a 12 GHz photodetector (PD 2), and digitized by a real-time oscilloscope (Keysight DSOS804A) with an effective bandwidth of 4 GHz, see Fig. \ref{fig1}(a). The averaged optical spectra are recorded by an optical spectrum analyzer (OSA, Agilent 86142B) with a resolution of  0.06 nm. The DFT path includes a 12 km SMF, corresponding to a group delay dispersion of $\sim$360 ps/nm around 1030 nm. 
The resolution of single-shot DFT spectra is estimated to be 0.58 nm \cite{goda2013dispersive}.

To initiate mode-locking in the MO, we inject seed pulses from a home-built all normal dispersion fiber laser \cite{chong2006all} into the MO. The seeding pulses has a $\sim$10 nm bandwidth that covers the filter separation. Filter 1 is set to 1030 nm while filter 2 is adjusted between 1034 nm and 1038 nm. When the single mode pump and multi-mode pump are adjusted to 400 mW and 1.1 W, respectively, stable single pulse mode-locking can be achieved by adjusting the waveplates. Once initiated, the mode-locked pulse circulates stably in the MO after blocking the seed laser. By increasing the filter separation to 7 nm and increasing the pump (multimode pump, the same below) power to 1.3 W after starting mode-locking \cite{chen2021starting}, we get a stable mode-locking state whose characteristics are depicted in Figs. \ref{fig1}(b-e). The pulse spectrum spans about 50 nm at $-$20 dB level and the mode-locked pulse train has a repetition rate of 11.81 MHz. The corresponding radio-frequency (RF) spectrum of the pulse train exhibits a signal-to-noise of $\sim$70 dB with a resolution bandwidth of 1 Hz (see Fig. \ref{fig1}(d)). 
The chirped pulse duration in this state is $\sim$4.6 ps as determined by an intensity autocorrelator (see Fig. \ref{fig1}(e)). 

\section{Observation of pulsating dissipative solitons}
\subsection{Single soliton and soliton molecule period-2 pulsation}

The laser tends to operate in the stable single pulse regime when the filter separation is relatively large. When decreasing the filter separation to 4 nm by tuning filter 2 to $\sim$1034 nm under a pump power of 1.1 W, pulsating DSs can be observed frequently in the MO.
Operation regime of the MO can be adjusted by the polarization dynamics. Various pulsation states can be accessed by rotating the waveplates under this relatively narrow filter separation. This also enables a comprehensive study of pulsation dynamics in our MO. 
We present two typical examples of single pulse period-2 pulsation states in Figs. \ref{fig2} (a-f). A pulse train is plotted in Fig. \ref{fig2}(a), showing that pulses in successive round trips have significant intensity change. Figure \ref{fig2}(b) shows the periodic change of the DFT measured single-shot spectra in 50 round trips; pulse energy derived by integrating the single-shot spectra power also exhibits period-doubling.
To show more details of the pulsation, spectra of two successive round trips (i.e., 1$^{st}$ and 2$^{nd}$ round trip) are shown in Fig. \ref{fig2}(c). 
These two spectra show variations in 3 dB bandwidth (changing between 5 and 11 nm), spectral fringe features and average power. 
The large spectral bandwidth breathing can be attributed to the different spectral broadening levels in the Mamyshev regenerators when the pulse energy changes in the pulsation. 
Note that the narrower spectrum still covers the two filters (4 nm separated). Thus, it can be regenerated in the cavity to retain mode-locking. 
By rotating the waveplates, the output pulse energy variation can be very large in some cases. Figures \ref{fig2}(d-f) show an extreme example, where one of the output DSs can be very weak. The output pulse energy change (defined as the ratio between the strong and the weak one) is as large as 40 times within one pulsation cycle by integrating the single-shot spectra (note that the time domain measurement does not show such a high ratio as the detector is saturated for the strong pulse in order to measure the weak one). 
This corresponds to the largest pulse energy variation during pulsation in mode-locked lasers, to our knowledge.
As a caveat, the PBS output port is a nonlinear output port and the intracavity pulse energy may not change as large as the output (see also simulations below). In other words, the large output energy variation between two successive round trips may arise from significant nonlinear polarization rotation in two round trips. This may help the intracavity pulse to retain an energy level required to keep consistent spectral broadening in the MO. Therefore, the large output energy variation may not provide direct experimental evidence for the extreme pulsation yet \cite{chang2015extreme}.

By increasing the pump power to 1.2 W, a pulsation state comprising two bounded DSs (i.e., a soliton molecule) can be obtained by rotating the waveplates. Figure \ref{fig2}(g) shows a pulse train measured without DFT, which indicates period-2 pulsation. The associated single-shot spectra are presented in Fig. \ref{fig2}(h). Stretching and compression of the spectra is evident and the weak pulse has an energy about half of the strong one. 

To compare the pulsating spectra, two consecutive round trips spectra are shown in Fig. \ref{fig2}(i). The measured spectra exhibit evident spectral modulation, which arises from the interference between two closely spaced DSs \cite{gui2018soliton}. The contrast of the spectral fringe is lower than unity due to the finite resolution of the DFT method. Similar to the single soliton pulsation in Fig. \ref{fig2}, the 3 dB bandwidth of the soliton molecule spectra change from 25 nm to 16 nm in the pulsation. 
Fourier transform of the single-shot spectra in Fig. \ref{fig2}(i) is plotted in Fig. \ref{fig2}(j). Peaks at non-zero time delay can be observed and they indicate that the 2 pulsating DSs are separated by 4.65 ps. This separation remains the same during pulsation, which differs from the vibrating solitons whose pulse separation changes periodically \cite{herink2017real,krupa2017real,Liu_PRL2018}. 

\subsection{Period-3 pulsation and soliton explosion}


For higher pump power (e.g., exceeding 1.2 W), multi-soliton state occurs frequently. 
We also observe pulsation in a two-DS state with the pulses widely spaced. Figure \ref{fig3}(a) plots the nonuniform pulse train in the time domain. This state shows period-3 pulsation instead of period-2 pulsation. 
The RF spectrum of the pulse train shows RF tones at the harmonics of 11.8 MHz/3 besides the fundamental 11.8 MHz (Fig. \ref{fig3}(b)), which further validates the period-3 pulsation of this state. 

Single-shot spectra over 6000 round trips are presented in Fig. \ref{fig3}(c). 
We first zoom in the data of 50 round trips around the 3000$^{th}$ round trip in Fig. \ref{fig3}(d) (rectangular shaped region in Fig. \ref{fig3}(c)). 
Although the two DSs both have period-3 pulsation, they pulsate in two ways. The single-shot spectra within 3 successive round trips are depicted in Fig. \ref{fig3}(e1-e3). DSs show three different spectral shapes marked as A, B and C. DS \textrm{I} pulsates in a route as A->B->C, while DS \textrm{II} pulsates as C->A->B. In other words, both DSs exhibit three spectral shapes in pulsation, but experience them in different orders. This also corroborates the pulse energy change delay in Fig. \ref{fig3}(d). Such a delayed spectral evolution can reduce the total pulse energy change than a synchronized spectral evolution. 

Besides the periodic pulsation, there are some round trips where the DSs experience soliton explosion (see ellipse regions in Fig. \ref{fig3}(c)).  These soliton explosions associated abrupt collapses occur randomly and details of the explosion are indicated by arrows in Figs. \ref{fig3}(f1-f4). 
Typically, DSs recover to a steady state in soliton explosion \cite{cundiff2002experimental,runge2015observation}, and soliton explosion in pulsating DSs has only been recently reported for a NPR mode-locked fiber laser \cite{peng2019experimental} and carbon-nanotube mode-locked fiber lasers \cite{chen2019dynamical,zhou2020breathing}. Here, soliton explosion for pulsating DSs is observed in a MO. Instead of collapsing into chaotic and wide spectra as shown in Ref. \cite{peng2019experimental}, the explosions in our case are mild in both time and frequency domain. We attribute this relatively mild pulsation to the strong shaping effect on pulses of Mamyshev regenerators; extreme explosion cannot survive in concatenated offset frequency filtering. The explosion lasts for about 5 round trips ($\sim$400 ns) which is much smaller than the $\sim$20 $\mu$s explosion time scale reported in ref. \cite{peng2019experimental}.
The pump power for this state (1.3 W) is higher than the mode-locking threshold (1.1 W) and we believe that strong nonlinearity is essential to the observation of explosion in pulsation in our MO. As an aside, when one of the DSs undergoes the explosion, the other one pulsates regularly, which suggests that the interaction is relatively weak for these two widely spaced DSs.  

\subsection{Long period and chaotic pulsation}

In addition to the short period pulsation shown above, pulsation with a relatively long period can also be possible in fiber lasers. For example, long-period pulsation (up to 170 round trips) associated with spectral bandwidth breathing has been reported in NPR mode-locked lasers \cite{peng2019breathing}. This type of long-period pulsation can also be observed by careful control of the polarization state in our MO under a pump power of 1.15 W. Figure \ref{fig4}(a) plots the pulse train of an observed period-9 pulsation. 
Figure \ref{fig4}(b) records the spectral evolution of 200 round trips, where we also notice abrupt spectral breathing during long-period pulsation. The spectrum widens about 1.5 times compared to the narrowest one (from 10 nm to 15 nm for 3 dB bandwidth), which creates significant spectral content in the wings of the spectrum. 
We plot the spectral evolution within one pulsation period in Fig. \ref{fig4}(c). The figure shows that the  gradually decreases and transfers to the spectral wings during pulsation and this evolution repeats in 9 round trips. Note that the periodicity is not very robust in this long period pulsation, which can be seen from the power variation in Fig. \ref{fig4}(b).


Beyond this long period pulsation, chaotic state can also be realized in our MO. The route towards the chaotic state can also be controlled by fine control of the waveplates (for example, QWP in Fig. \ref{fig1}(a)). 
When the angle of the quarter waveplate is about 165$^\circ$, the MO operated in the stable single 
pulse state (see Fig. \ref{fig4}(d)). The pulse energy decreases as the waveplate rotates counterclockwise. Period-2 pulsation appears at a angle of 153$^\circ$. Further tuning of the quarter waveplates leads to a chaotic state (Fig. \ref{fig4}(e)). The pulse train can no longer repeat itself in this state. 
Further experiments show that all soliton forms (single soliton, multi-soliton and soliton molecule) can bifurcate from stable operation to multi-period pulsating and even to chaos, which suggests that chaotic pulsation is a ubiquitous phenomenon in our MO.




\section{Simulation of pulsation in Mamyshev oscillators}

\subsection{Period-2 pulsation in simulations}
To have a deeper understanding of our experimental observations, we use a lumped model to simulate the pulsating DS dynamics in the MO. 
Details of the simulations are provided in Appendix, which also includes parameters used in generating the pulsation dynamics. Since polarization dynamics is important in our experiment, we use a vector generalized nonlinear Schr${\rm \Ddot{o}}$dinger equation (NLSE) to simulate pulse propagation in the fibers and Jones matrix to model the waveplates and PBS. Consistent with experiments, small filter separation can lead to pulsation in simulations. Figure \ref{fig5}(a) shows an example of generating pulsating DSs by reducing the filter separation from 6 nm to 3 nm from a stable mode-locking state. In agreement with the experiment, the pulsating DSs are also associated with spectral breathing, with the 3 dB bandwidth varying from 10 nm to 12 nm in the pulsation (see inset in Fig. \ref{fig5}(a)). In another case, pulsation can also transition from stable mode-locking by adjusting the waveplates with the filter separation and bandwidth being 3 nm and 0.8 nm, respectively (see Fig. \ref{fig5}(b)).

The pulse energy change during pulsation can be adjusted by increasing gain of the active fibers and rotating the waveplates in simulations (see Appendix). Figures \ref{fig5}(c, d) show a pulsation state with a relatively large pulse energy change and the corresponding normalized spectra. The corresponding temporal and spectral dynamics within two successive round trips are shown in Figs. \ref{fig5}(e, f). 
Pulsating DSs have slightly different peak powers at the same position in these two round trips (see white line in Fig. \ref{fig5}(e)), resulting in different spectral broadening in Fig. \ref{fig5}(f). The insets of Fig. \ref{fig5}(f) show the spectra right after filter 1 whose relative spectral power can be referred to the colorbar. 
The simulated pulse energy change is 3.3 times in Figs. \ref{fig5}(c-f). Although this is still much smaller than the 40 times observed in Fig. \ref{fig2}(e),  we believe a larger energy change can be possible by further exploration of the parameters. Simulation shows that polarization effects and using PBS as the output port are important to have a relatively large pulse energy change in pulsation. As a comparison, a typical pulsation state simulated using a scalar model (see Appendix) only have a pulse energy change of $\sim$1.006 times (Fig. \ref{fig5}(g)). 
As noted in Section 3A, the intracavity pulse energy change may be smaller than the PBS output. 
This is verified in the simulation; the intracavity pulse energy only changes 1.3 times in the state of Fig. \ref{fig5}(c). As an aside, QWP can be rotated over 360$^\circ$ in some filter and pump power settings without losing mode-locking, which is also consistent with the observation noted in Section 2. 


Pulsating soliton molecule can also be simulated by increasing the gain saturation energy $E_{\rm sat}$ (corresponding to increasing pump power in experiments, see Appendix), as shown in Fig. \ref{fig5}(h). Single-shot spectra in two consecutive round trips (Fig. \ref{fig5}(i)) and their zoom-in (Fig. \ref{fig5}(j)) show an imperceptible change of spectral structure along with a mild energy change. Figure \ref{fig5}(k) is the Fourier transform of the simulated spectra, indicating that the pulse separation remains unchanged in this state. 

\subsection{Explosion and chaotic pulsation in simulations}
Period-3 pulsation and DS explosion can also be attained by increasing the small signal gain $g_0$ as well as saturation energy $E_{\rm sat}$ and finely tuning the waveplates (see Appendix). Figures \ref{fig6}(a-c), (b), and (c) show the spectral evolution over 2000 round trips, over 50 round trips and single-shot spectra within a pulsation period, respectively. Note that our simulation time window is limited to 100 ps (only one DS included in the window) so as to reduce the computation complexity. This simplification would not impact the pulsation dynamics significantly, since the interaction between 2 DSs in Fig. \ref{fig3} is negligible. Similar to experimental observation, the simulated period-3 pulsation is associated with soliton explosion (indicated by arrows in Fig. \ref{fig6}(a)). Zoom-in of the explosion is shown in Figs. \ref{fig6}(d, e). The spectrum collapses into a distorted one (around the 922 $^{nd}$ round trip) and the pulse energy changes irregularly during the explosion (white line in Fig. \ref{fig6}(d)). The DS recovers to the regular pulsation state after about 5 round trips. This explosion time scale is close to the measurements in Fig. \ref{fig3}. Figure \ref{fig6}(e) shows another example of soliton explosion during pulsation where the explosion lasts slightly longer ($\sim$30 round trips). 
Figures \ref{fig6}(f1, f2) show the simulated temporal shift during explosion shown in Figs. \ref{fig6}(d, e), respectively. Such a temporal shift during explosion was reported in refs. \cite{runge2015observation,peng2019experimental}. And it may arise from the spectral center frequency and group velocity change during explosion (power variation during explosion can lead to different spectral broadening in the MO). Note that the period-3 pulsation can be obtained under the similar conditions as used in Fig. \ref{fig5}, while DS explosion can only be observed by increasing $g_0$ and $E_{\rm sat}$ to relatively large values (see Table \ref{tab2}). Therefore, we believe that high pump power is essential to the emergence of soliton explosion in MOs, which is consistent with the experimental observations. And it is in contrast to the explosion observed in ref. \cite{peng2019experimental}, which occurs at a relatively low pump power.

Long period and chaotic pulsation can also be simulated by rotating the waveplates (see Appendix for used parameters). Figure \ref{fig6}(g) shows the spectral dynamics for a period-14 pulsation, while Fig. \ref{fig6}(h) shows three normalized representative spectra in a pulsation period (the 1$^{st}$ spectrum corresponds to the narrowest spectra in Fig. \ref{fig6}(g)). 
Further rotation of the waveplates can generate chaotic pulsation as shown in Fig. \ref{fig6}(i). The corresponding pulse energy change loses periodicity. These simulations are in qualitative agreement with the observation in Fig. \ref{fig4}. 


\section{Discussion and conclusion}

The spectral regeneration and filtering in MOs make them distinct from other mode-locked lasers. Our measurements unveil that short-period, long-period and chaotic pulsations all exist in MOs. Under relatively high pump powers, DSs can even explode in pulsation. Numerical simulation verifies the existence of these dynamics in MOs. 
Our observation confirms that pulsating DSs are an eigen-solution of the master equation governing the DS dynamics in MOs. Since nonlinear gain/loss effects (e.g., quintic dissipative terms in CGLE) are essential for the generation of pulsation in mode-locked lasers \cite{akhmediev2001pulsating}, the saturable absorption based on the Mamyshev regenerator may provide these effects (slow gain dynamics are not included in our simulation). However, NPR-like effects may also contribute to needed nonlinear gain/loss terms and impact the observed dynamics. It can be interesting to build a polarization maintaining MO to eliminate the polarization dynamics and study the scalar pulsation dynamics in MOs. Combining DFT with the time lens method \cite{ryczkowski2018real} may reveal more pulsation dynamics in MOs. Such future work can be helpful to further reveal the role of Mamyshev regenerator in pulsation, experimentally and theoretically. Finally, it can contribute to a better understanding of complex pulsation dynamics including the DS explosion in pulsation. Moreover, our work is based on an all-normal-dispersion MO and it can be interesting to explore the pulsation dynamics in MOs with other dispersion configurations.

In conclusion, we have experimentally and numerically observed pulsating DSs in a MO. These pulsating DSs arise under a relatively high pump powers and small filter separation and bandwidths. We observe various pulsation dynamics for mode-locking states comprising a single DS or 2 DSs. By controlling the polarization state, dramatic pulse energy change with output pulse energy change up to 40 times is possible. Soliton explosion during pulsation has also been observed in the MO under a relatively high pump power. 
Chaotic pulsation state also exists in the MO. Simulations are in qualitative agreement with the observations.  
Our results not only show that narrow filter separation should be avoided for stable operation of high power MOs but also add to the pulsation dynamics in dissipative optical systems.

\section*{Appendix}

\begin{table*}[t]      
\centering
\caption{\bf Parameters of the optical fibers}
\label{tab1}
    \begin{tabular}{ccccccc}    
        \hline           
        Fiber & Length (m) & $\beta_{2}$ (ps$^2$ km$^{-1}$) & $\beta_{3}$ (ps$^3$ km$^{-1}$ ) & $\gamma$ (W$^{-1}$ km$^{-1}$) & $L_B$ (m) & Gain bandwidth (nm)\\   
        \hline
        YDF1 & 3.1&24.9& 59& 4.8 & 1 & 30\\       
        YDF2 & 3.1&24.9& 59& 4.8 & 1 & 30\\ 
        SMF1 & 0.7&22.2& 63.8& 4.4 & 1 & NA\\
        SMF2 & 2.1&22.2& 63.8& 4.4 & 1 & NA\\
        \hline
        \end{tabular}
\end{table*}

\begin{table*}[t]      
\centering
\caption{\bf Other parameters used in the simulations}
\label{tab2}
    \begin{tabular}{ccccccc}    
        \hline          
        Figure & Filter separation (nm) & Filter bandwidth (nm) & $g_0 \rm (m^{-1})$ & $E_{\rm sat} \rm (nJ)$  & $\phi_{\rm QWP} (^\circ)$ & $\phi_{\rm HWP1} (^\circ) $\\  
        \hline
        Fig. 5(a) & 6 to 3 & 0.8 & 10 & 1 & 90 & 20 \\
        Fig. 5(b) & 3 & 0.8 & 9 & 1 & 90 to 50 & 50 \\
        Fig. 5(c) & 3 & 0.7 & 13 & 1.1 & 83 & 23 \\
        Fig. 5(g) & 2.5 & 1 & 10 & 1 &  NA & NA   \\
        Fig. 5(k) & 4 & 0.9 & 5.8 & 12.2 & 19.1 & 20 \\
        Fig. 6(a) & 3 & 0.7 & 31.3 & 3.8 & 89.86 & 20 \\
        Fig. 6(g) & 2.5 & 0.8 & 10 & 1 & 90 & 26 \\
        Fig. 6(i) & 2.5 & 0.8 & 10 & 1 & 90 & 24 \\
        \hline
        \end{tabular}
\end{table*}

Simulations is based on a lumped model. In this model, pulse propagation in the fibers is simulated by the coupled generalized nonlinear Schr${\rm \Ddot{o}}$dinger equation (NLSE) \cite{PhysRevLett.101.153904}:
\begin{eqnarray}
\frac{\partial u}{\partial z}=  i \beta u-\delta \frac{\partial u}{\partial t}-\frac{i \beta_{2}}{2} \frac{\partial^{2} u}{\partial t^{2}}+i \gamma\left(|u|^{2}+\frac{2}{3}|v|^{2}\right) u+ \nonumber\\\frac{i \gamma}{3} v^{2} u^{*}+\frac{g}{2} u+\frac{g}{2 \Omega_{g}^{2}} \frac{\partial^{2} u}{\partial t^{2}} \\
\frac{\partial v}{\partial z}=-i \beta v+\delta \frac{\partial v}{\partial t}-\frac{i \beta_{2}}{2} \frac{\partial^{2} v}{\partial t^{2}}+i \gamma\left(|v|^{2}+\frac{2}{3}|u|^{2}\right) v+\nonumber\\\frac{i \gamma}{3} u^{2} v^{*}+\frac{g}{2} v+\frac{g}{2 \Omega_{g}^{2}} \frac{\partial^{2} v}{\partial t^{2}}
\end{eqnarray}
where $u$ and $v$ are the slowly varying pulse envelopes of fast and slow axes of the cavity, respectively. $2 \beta=2 \pi/(\lambda \Delta n)=2 \pi/L_{B}$ and $2\delta=\beta\lambda/(\pi c)$ are the wave number difference and inverse group velocity difference, respectively ($L_B$ is the beat length). $\beta_2$ is the group velocity dispersion, $\gamma$ is the Kerr nonlinearity coefficient, $g=g_{0} \exp \left[-\int\left(|u|^{2}+|v|^{2}\right) dt / E_{\rm sat}\right]$ is the gain, with  $g_0$ and $E_{\rm sat}$ being the small signal gain and gain saturation energy, respectively; $\Omega_g$ is the gain bandwidth. For passive fibers, small signal gain is chosen as $g_0$=0 m$^{-1}$. The parameters of the fibers are given in Table \ref{tab1}. The fiber parameters are fixed for all simulations.

Simulations start from two weak pulses along the fast and slow axes of the cavity (a weak pulse for the scalar model) to speed up the convergence of our simulations. Starting from the Yb-doped gain fiber (YDF1), the weak pulses are amplified and experience spectral broadening due to SPM. The spectrum further broadens in the passive fiber (SMF1) before filtered at both axes by filter 1 with a gaussian profile. The filtered pulse experiences amplification and spectral broadening for the second time in YDF2 and SMF2. Then the pulse passes a QWP, a PBS and two HWPs, which are modeled by the Jones matrixes (angles are defined with respect to the x-axis, which is assumed as one of fiber eigen-polarization axes). The PBS is inserted before the second HWP (HWP2) to output the x-polarized component. Finally, the pulse is filtered by  filter 2, whose transmission window is shifted to longer wavelengths from filter 1. Then the loop is closed. HWP 2 is fixed to have an orientation of 90$^\circ$. The parameters used to generate the various pulsation dynamics are listed in Table \ref{tab2}. As for the scalar model used in Fig. \ref{fig5}(g), we used a model similar to ref. \cite{wang2019pattern}. We neglect the polarization effect of all components and get the result from a linear output port with a coupling ratio of 20 $\%$ after the SMF2.

The parameters used in simulations are in reasonable agreement with experimental measurements. Some conditions may differ from the experiments; for example, the active fibers in two arms share the same gain parameters ($g_0, E_{\rm sat}$) and the filter properties are adjusted in simulations (fixed in experiments). These slight discrepancies arise from the difficulty in modeling the complex MO accurately and the need to reduce the parameter space to explore and computation complexity. However, the discrepancies will not cause the loss of generality. For example, the simulated pulsating DS and other results we discussed above can be attained under different lengths of YDF and SMF by adjusting $g_0$ and $E_{\rm sat}$ as well as filter parameters.


\begin{acknowledgments}
This work is supported by the National Natural Science Foundation of China (NSFC) (51527901, 61575106), and fundamental Research Funds for the Central Universities (BUPT 2021RC08). The authors declare no conflicts of interest. Data underlying the results presented in this paper are not publicly available at this time but may be obtained from the authors upon reasonable request.
\dots.
\end{acknowledgments}


\nocite{*}

\bibliography{apssamp}

\end{document}